# Towards a Small Prototype Planet Finding Interferometer (SPPFI) – The next step in planet finding and characterization in the infrared

W.C. Danchi[1], D. Deming[1], K. G. Carpenter[1], R. K. Barry[1], P. Hinz[2], K. J. Johnston[3], P. Lawson[4], O. Lay[4], J. D. Monnier[5], L. J. Richardson[6], S. Rinehart[1], W. Traub[4]

[1]NASA Goddard Space Flight Center, Astrophysics Science Division, 8800 Greenbelt Road, Greenbelt, MD 20771
[2]University of Arizona, Department of Astronomy, Tucson, AZ
[3]Naval Observatory, Washington, DC
[4]Jet Propulsion Laboratory, California Institute of Technology Pasadena, CA
[5]University of Michigan, Department of Astronomy, Ann Arbor, MI
[6]American Astronomical Society, Washington, DC

**ABSTRACT**

During the last few years, considerable effort has been directed towards large-scale (>> $1 Billion US) missions to detect and characterize earth-like planets around nearby stars, such as the Terrestrial Planet Finder Interferometer (TPF-I) and Darwin missions. However, technological and budgetary issues as well as shifting science priorities will likely prevent these missions from entering Phase A until the next decade. The secondary eclipse technique using the Spitzer Space Telescope has been used to directly measure the temperature and emission spectrum of extrasolar planets. However, only a small fraction of known extrasolar planets are in transiting orbits. Thus, a simplified nulling interferometer, which produces an artificial eclipse or occultation, and operates in the near- to mid-infrared (e.g. ~ 3 – 8 or 10 µm), can characterize the atmospheres of this much larger sample of the known but non-transiting exoplanets. Many other scientific problems can be addressed with a system like this, including imaging debris disks, active galactic nuclei, and low mass companions around nearby stars. We discuss the rationale for a probe-scale mission in the $600-800 Million range, which we name here as the Small Prototype Planet Finding Interferometer (SPPFI), pronounced "spiffy".

**1. INTRODUCTION**

Over the course of the last 10 years, more than 215 extra solar planets have been discovered using a variety of observational methods, including precision radial velocity techniques (Mayor & Queloz 1995, Marcy & Butler 1998), transit searches (Charbonneau et al. 1999, Alonso et al. 2004), microlensing (Bennett et al. 1999), imaging (Chauvin et al. 2005), and pulsar timing (Wolszczan et al. 1992). A summary of data on known exoplanets can be obtained on the website (Schneider 2007). Of the 215 presently known planets, the vast majority of them (203) were discovered using the radial velocity (RV) technique, and as the technique has improved from a precision of ~3 ms$^{-1}$ from a few years ago, to about 1 ms$^{-1}$ currently. The lower limit of detectable masses has gone from around a Saturn mass (~0.3 $M_J$) to Neptune (0.054 $M_J$) or Uranus (0.046 $M_J$) masses.

The long term focus of planet finding efforts at NASA has been directed towards the search for extrasolar terrestrial planets, like those in our own solar system, within the habitable zone around stars similar to our own Sun, e.g., F, G, and K main sequence stars. At the present time, budgetary considerations within NASA have put the Terrestrial Planet Finder Interferometer (TPF-I) mission beyond the foreseeable future.

At the present time, it is possible to build a small infrared interferometer, with apertures of the order of 1 m diameter and passive cooling to about 60K, that can make significant progress towards characterizing the atmospheres of a large number of known exoplanets, as well as measuring the debris disks around candidate stars for TPF-I and Darwin (building on recent progress from the Spitzer Space Telescope, and SCUBA [Bryden et al. 2005, Beichman et al. 2005, Greaves et al. 1998, Holland et al. 1998]).

In this paper we focus on the exoplanet parameter space related to TPF and Darwin that can be explored with a modest infrared interferometer system, herein called SPPFI. The SPPFI mission concept is closely related to two mission concepts that have been discussed extensively in the past. They are the Fourier Kelvin Stellar Interferometer (FKSI) (Danchi et al. 2003a), pronounce "foxy", and the Pegase mission concept (Duigou et al. 2006). Some of the graphical elements in this paper are taken from the former concept. Our purpose in this paper is not to endorse a specific mission concept, only to illustrate what science can be done with a modest system. The remainder of this paper is focused on the case for the two main science goals for a modest interferometer – measuring the physical characteristics of known exoplanets (§ 2) and debris disks (§ 3) – and also on the required instrumental characteristics and performance (§ 4).

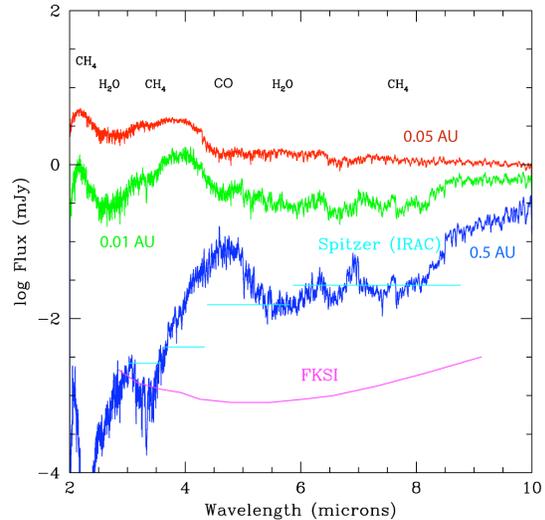

**Figure 1**. (a) Left panel. Characteristics of exoplanets that can be measured using a modest infrared space interferometer. (b) Right panel. Sample theoretical spectra of exoplanets with different semi-major axes compared to Spitzer IRAC sensitivity, and that for a sample interferometer, the FKSI system.

## 2. EXOPLANETS

**Some characteristics of known extrasolar planets.** The currently known planets cover an impressively broad range of semi-major axes from about 0.02 AU up to about 6 AU and masses from about 10 $M_J$ to about 0.02 $M_J$. These planets also cover a very wide range of eccentricities, which are quite different and distinct from our own Solar System.

The origin of the close-in hot Jupiter-like planets with very short periods is being actively investigated by Gaudi et al. (2005) as is the much larger than anticipated eccentricity distribution.
Of the 215 planets, currently eleven (14) are transiting planets, and for these much more physical knowledge can be gained since the inclination angle (i), the planet radius ($R_P$) and hence the mean density can be determined. Precision photometry has been used with the transiting planets to determine the temperature of the atmosphere [Deming et al. (2005, 2006), Charbonneau et al. (2005)], chemistry [Charbonneau et al. 2002, Vidjal-Madjar et al. 2003], and albedo [Rowe et al. 2006]. Recently, the first emission spectra of transiting planets have been published by Richardson et al. (2007) for HD 209458b and Grillmair et al. (2007) for HD 189733b, using the secondary transit observed with Spizer Space Telescope.

The frequency of extrasolar planets increases rapidly as the mass decreases with an approximately 1/M distribution and it is expected that the current pace of detected planets (of the order of about 30 per year) will continue for the foreseeable future.

In the future, we need to focus on studies of the physical characteristics of the known planets as well as to continue observational programs to detect lower mass planets and more planetary systems. The wide range of masses and orbital characteristics provides for a laboratory of exoplanet astrophysics that will allow us to place our own Solar System and those like our own in the larger context of stellar and planetary origin and evolution.

Figure 1(a) displays a table of planetary characteristics and what can be determined using a modest nulling interferometer, such as planetary temperature, radius, mass density, albedo, surface gravity, and atmospheric composition, including the presence of water. To date, progress has been made on the physical characteristics of planets largely through transiting systems, but a small planet finding interferometer such as SPIFFY can measure the emission spectra of a large number of the non-transiting ones.

**Parameter space to be explored with a SPPFI system.** Transiting planets have been extraordinarily useful in terms of expanding and deepening our knowledge of the physical parameters of extrasolar planets. By careful analysis of transit light curves it is possible to fit the planet radius and inclination angle to a fairly high precision, allowing a measurement of the mean density of the planets. Most of the planets lie near the density of water, which is 1 g cm$^{-3}$. None of the transiting planets is considered "rocky", for example, the mean density of the Earth is ~ 5 g cm$^{-3}$, which is far larger than water.

Many molecular species, such as carbon monoxide, methane, and water vapor, have strong spectral features in the 3-8 µm region, as can be seen in Figure 1(b), which displays model atmospheres for extrasolar planets calculated by Seager (2005), for planets at various distances from the host star. The red curve shows the theoretical spectrum of a very hot, close-in planet at 0.05 AU, while the blue curve displays spectrum for a much cooler planet ten times further out, at 0.5 AU. Also displayed are sensitivity curves for the IRAC instrument on Spitzer (light blue) and FKSI (purple). Clearly such a mission concept has sufficient sensitivity to detect and characterize a broad range of extrasolar planets. If the telescopes are somewhat larger than has been discussed in some of the exisiting mission concepts (e.g., 1-2 m) and are somewhat cooler (e.g., < 60K) so that the interferometer can operate at longer wavelengths, it is possible for the SPPFI system to detect earth-like planets around the nearest stars. This is especially important now that there is an increasing belief that lower mass planets are very common, based on the detection of the 5.5 Earth mass planet using the microlensing approach (e.g. Beaulieu et al. 2006, Gould et al. 2006).

## 3. DEBRIS DISKS

**Recent progress on debris disks.** During the past few years, observers using the Spitzer Space Telescope and the James Clerk Maxwell Telescope have made progress on studies of the frequency of debris disks around solar type stars, and the characteristics of debris disks previously discovered by IRAS (Auman et al. 1984, Plets & Vynckier 1999) and ISO (Spangler et al. 2001). The major focus of such studies has been on the decay of planetary debris disks (e.g., Rieke et al. 2005) and on the frequency of such disks around solar type stars (Bryden et al. 2006, Beichman et al. 2005, Kim et al. 2005). Debris disks have been discovered around stars having a range of masses, such as from A to K stars (e.g. Beichman et al. 2005), and white dwarf stars as well (Reach et al. 2005).

The lifetimes of debris disks of a large sample of A stars (masses around 2.5 M$_\odot$) have been shown to be ~ 150 Million years or less, with a substantial population of stars much older than that with excesses, presumably caused by relatively recent episodes of collisions of large planetesimals (Rieke et al. 2005).

The Spitzer MIPS observations at 70 µm of Beichman et al. (2005) found that there were 6 of 26 main sequence stars (with RV known planets) with substantial debris disks, and hence that excess emission was as common around the main sequence stars as earlier observations of A and F stars with IRAS. Another import thrust of research on debris disks with Spitzer has been to push down the detection threshold as measured by $L_{dust}/L_{star}$, from about 10$^{-3}$ with IRAS to as low as about 10$^{-5}$, as has been done by Kim et al. (2005) using Spitzer.

The morphology of debris disks has also been investigated around a number of A stars, including ε Eri, β Pic, Fomalhaut, and Vega (Holland et al. 1998, 2003) at 450 and 850 µm, and HR 4796A (Koerner et al. 1998, Jayawardhana et al. 1998) at 10 and 20 µm. These observations show a number of complex features, such as cavities, asymmetries, and clumps, which could be due to the resonant trapping of dust associated with a large planet, and other important dynamical phenomena such as collisional cascades of dust from interactions between planetesimals, and the interplay with effects like radiation pressure and Poynting-Robertson and gas collisional drag.

These studies have focused on the cold outer regions of the debris disk systems, where the dust temperatures are around 50 K, which is a different population of dust than is most important for the search for earth-like planets in the habitable zone, where the dust should be around 300 K. Present limits are several orders of magnitude greater than expected for the hotter material in our own zodiacal cloud (Dermott et al. 2002), which about $L_{dust}/L_{star} \sim 10^{-7} - 10^{-8}$. Cooler material in our Kuiper belts is predicted to be of the order of $10^{-6} - 10^{-7}$ (Stern 1996).

**Disk parameter space to be explored with a SPPFI system.** Although a great deal of progress has been made in recent years on debris disk, a variety of instrumental limitations will prevent the community from achieving the progress that we need for the TPF and Darwin mission. There are two types of limitations. For Spitzer observations and likewise for Herschel and JWST, the main problem is that dust excesses are measured relative to the stellar spectrum, which is extrapolated to the infrared and compared with the observations. Calibration uncertainties limit the excesses measured to a few percent over the estimated stellar spectrum. This gives the limit from Spitzer of about 1000 Solar System Zodis (SSZs). Another problem with the current observations is that they are most sensitive at the longer wavelengths, namely, 70 µm for Spitzer and in the submillimeter for the SCUBA observations at JCMT. Thus present observations constrain extrasolar zodiacal emission from the population of dust associated with distances of 10 AU and greater, which may have very different characteristics than the population of dust at 1 AU near the habitable zone, of most interest to TPF-I and Darwin.

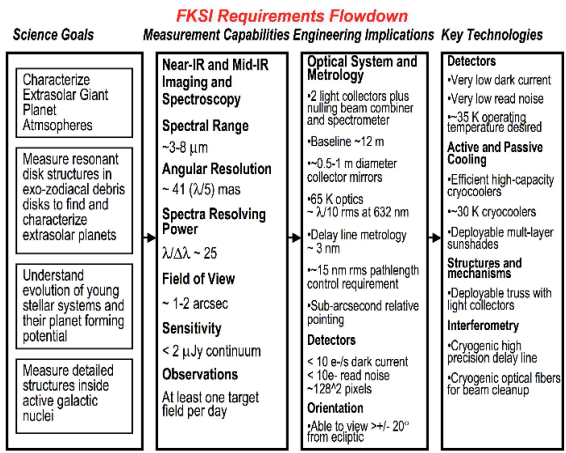

**Figure 2**. Example of a requirements flowdown for a sample small interferometer mission.

Ground-based nulling interferometers such as the Keck Interferometer Nuller and the nulling instrument on the Large Binocular Telescope Interferometer (LBTI), will make significant progress on this latter population of dust since they will observe at 10 µm near the peak of the 300 K blackbody emission. However, our present understanding of the limitations of ground-based warm nulling systems is that they will push the limit to about 100-150 SSZs around nearby stars, still far larger than is needed for sizing the TPF-I and Darwin missions. The reason ground based nulling interferometers can make such significant progress is that they suppress the star light and enable a clearer distinction between stellar and zodiacal light, thus bypassing the limitations of a purely spectroscopic approach.

Another limitation to present studies is the lack of spatial resolution. This is especially significant as clumps and asymmetries in the distribution of debris disk material can be used to search for unseen planets. Furthermore it is vitally important to know that the emission is coming from the region of the habitable zone, and to know the amount of dust for the TPF target stars. It is unlikely that we can know for certain what are the best TPF target stars (from the standpoint of exozodiacal dust) without such a mission.

## 4. INSTRUMENTAL CONSIDERATIONS

**Requirements Flowdown & Sensitivity.** In order to determine the performance characteristics of an instrument or mission, we must demonstrate how the science requirements flow down into the instrumental characteristics. This has been done for a mission with similar science requirements, and is shown in Figure 2.

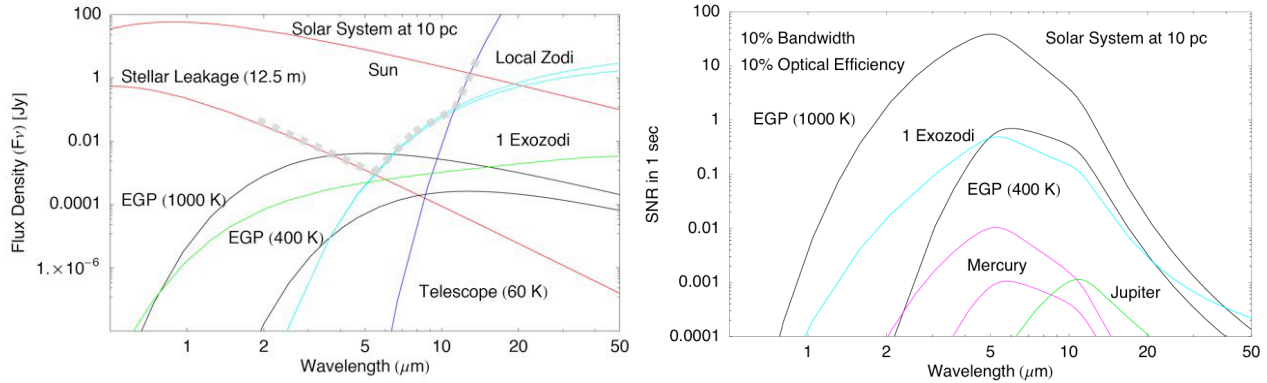

**Figure 3**. (a) Left panel. Flux density versus wavelength for a variety of sources affecting the performance of a cooled nulling space interferometer. (b) Right panel. Estimate of system performance (signal-to-noise ratio) for 1 m$^2$ collecting area, for 10% bandwidth and 10% optical efficiency.

We turn now to some recent calculations of signal and noise sources affecting the detection and characterization of exoplanets in the 3-8 µm region of the spectrum. Figure 3 (a) displays a calculation of the expected signal levels in Jy ($10^{-26}$ W m$^{-2}$ Hz$^{-1}$) as a function of wavelength (µm) for our Sun and viewed from a reference distance of 10 pc. We also display flux density levels for extrasolar giant planets at 400 K and 1000 K, for an exoplanet with radius (1.35 R$_J$) and albedo of HD209458b. Figure 3(a) also displays expected noise sources that affect the detection and characterization of exoplanets using a nulling interferometer including emission from the stellar leakage (due to the finite size of the star), emission from the exozodiacal dust cloud (based on the model of Reach et al. (2003)) from a dust cloud having the same characteristics as our own zodiacal cloud, emission from the local zodiacal cloud itself, and the thermal emission from the telescope system, assumed to be cooled to 60 K. Also drawn is a gray dashed line that shows the boundary created by the dominant noise sources as a function of wavelength, from stellar leakage, which is most important at short wavelengths, to the emission from the local zodiacal cloud, which dominates at the middle wavelengths, and the emission from the telescope itself at the longest wavelengths. We see that the region from 3 – 8 µm or even 10 µm is very favorable from a signal-to-noise ratio standpoint.

Figure 3(b) shows the expected signal to noise ratio in 1 s for a 1 m$^2$ mirror, assuming a 10% bandwidth, and 10% optical efficiency. Radii and albedos of the solar system planets were taken from the tables from Lissauer and DePater (2001). Clearly the warm and hot exo-Jupiters are easy to detect with this short integration time as are exozodiacal clouds. Given longer integration times, small hot planets such as Mercury are detectable, as are cooler, large planets like Jupiter.

**Angular resolution**. The distribution of distances of the currently known extrasolar planets peaks around 30-40 pc, which means the angular resolution required to resolve a planet from the star is in a range that is amenable to modestly sized interferometers in the near-infrared.

At 10 pc, Earth at 1 AU corresponds to an angular separation of ~0.1 arcsec, while Jupiter would be at about 0.5 arcsec for the same distance. This means that even a very modest system can probe spatial dimensions within the habitable zone of nearby stars. A reasonable number of planets have semi-major axes with apparent separations greater than this angular size, but a more detailed analysis by Danchi et al. (2003a,b) demonstrates that it is possible to detect and characterize planets even with semi-major axes much smaller than this. The angular scale needed to resolve the planet from the star is determined from the semi-major axis length in AU, scaled by the distance, and can be used to determine the required resolution of an imaging system, whether it be an interferometer or a filled-aperture telescope. For example, the nominal resolution ($\lambda/2B$) of a space interferometer with baseline, $B = 12.5$ m, at a center wavelength, $\lambda = 5$ µm, is ~ 0.04 arcsec. As a comparison, coronagraphic observations on JWST have an inner working angle (IWA) of ~ $4 \lambda / D$ ~ 0.7 arcsec at $\lambda=5$ µm and $D=6$ m. This much larger IWA means that JWST will not be able to probe the dust and planets in the region near the habitable zone for nearby stars, only zones corresponding to the outer planets and Kuiper belts.

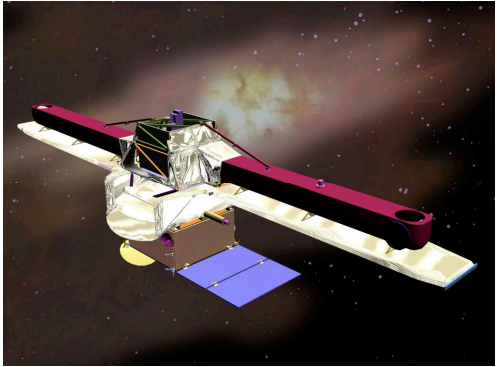

**Figure 4**. Artist's conception of a sample small prototype planet finding mission, that of the FKSI mission concept, which has been extensively studied.

**Spectral resolution.** The required spectral resolution requirement depends on the spectral width of the features that must be resolved. From Fig. 1(b), it is clear that modest spectral resolution, of the order of R=25-50, is adequate to resolve the molecular features and thereby measure the composition of the planet's atmosphere. In addition to mass, it will be important to observe the variation in the planet's infrared spectrum as a function of phase, and facilitate characterization of the planet's thermal state and composition. In favorable cases, inferences about dynamical properties of the atmospheres such as winds could be made, as discussed in Showman & Guillot (2002). This approach of exploiting the infrared brightness of exoplanets builds on the transit and radial velocity studies.

An additional benefit of an infrared stellar interferometer is the possibility of operation as a Fourier transform spectrometer, using the Double Fourier technique discussed by Marriotti & Ridgway (1988). In this case the major limitation to spectral resolution would be the length of the stroke of the delay line. A modest stroke length of ~ 2 cm would yield spectral resolutions of the order of 5000. Depending on the configuration of the detection systems, wide field imaging modes are also possible (e.g., Leisawitz et al. 2004).

**Number of accessible targets.** Given the collecting area of such an interferometer is of the order of a square meter, it is expected that it will be significantly more sensitive in the near-infrared and mid-infrared than the Spitzer Space Telescope, which has a primary mirror 85 cm in diameter. For 2 1-m diameter mirrors, the increase in collecting area is about a factor of 3, while if the mirrors are 2-m in diameter, the increase in collecting area is a factor of 10. Improvements in detector technology are also expected to yield further increases in sensitivity.

These considerations mean that the number of targets that can be observed are at least as many as can be observed with Spitzer, only with the provision that a passively cooled system has a field of regard that is restricted by the sunshade design. Studies to date indicate the field of regard can be at least as great as +/- 45 degrees from the ecliptic, allowing essentially all such sources to be observed, clearly indicating that many thousands of targets are accessible.

## 5. BRIEF REVIEW OF SOME EXISTING MISSION CONCEPTS

Two mission concepts that are related to the SPPFI concept have undergone significant development in the past five years. One called the Fourier-Kelvin Stellar Interferometer (FKSI), is a near- to mid- infrared nulling and imaging interferometer comprised of two 0.5 m diameter telescopes operating from 3 to 8 µm with a 12.5 m baseline on a boom, and passively cooled to about 60 K (see Fig. 4). The designs for FKSI are well advanced beyond what is normally done in pre Phase A (e.g., see Barry et al. (2005, 2006) and Frey et al. (2006)). Detailed cost estimates (grass roots, PRICE-H, and RAO) have been made and a mission like FKSI could easily fit into an Origins Probe level budget (~$600 Million) or New Horizons budget (~$750 Million). The other concept is called Pegase and was created in response to a call from CNES for free-flying spacecraft mission concepts (see Duigou et al. 2006 and references therein). These two concepts are similar and it is possible to make a joint mission from them, for example with mini-satellites attached to the ends of the FKSI booms. A six-month study would be of great benefit to refine the mission concept and reduce cost uncertainty and risk.

# 6. SUMMARY

We presented the rationale for a mission to characterize the atmospheres of presently known extrasolar planets and to measure emission from debris disks in the habitable zone in preparation for TPF. We explained why the 3-8 µm region is particularly favorable from a signal-to-noise ratio standpoint, and that such a mission can be accomplished with a relatively simple, passively cooled (60K) two-telescope space interferometer operating at L2. Such an interferometer only needs a collecting area ~ 1 $m^2$ in order to be able to detect and characterize a large number of existing extrasolar planets and debris disks. This unique mission is a desirable step in the path towards TPF. Technologies are well in hand from investments already made for TPF-I. Transition to Phase A could occur as soon as funding is available.